\documentclass[%
 reprint,
superscriptaddress,
prl,
]{revtex4-2}

\usepackage{graphicx}
\usepackage{dcolumn}
\usepackage{bm}
\usepackage{amsmath}
\usepackage{siunitx}
\usepackage[english]{babel}
\usepackage{amsfonts}

\usepackage{hyperref}
\hypersetup{
    colorlinks=true,
    linkcolor=blue,
    filecolor=blue,      
    urlcolor=blue,
    anchorcolor=blue,
    citecolor=blue
}

\DeclareMathOperator{\sinc}{sinc}

\begin{document}

\preprint{APS/123-QED}

\title{GeV-scale accelerators driven by plasma-modulated pulses from kilohertz lasers}

\author{O. Jakobsson}%
\affiliation{John Adams Institute for Accelerator Science and Department of Physics,University of Oxford, Denys Wilkinson Building, Keble Road, Oxford OX1 3RH, United Kingdom}%
\author{S. M. Hooker} 
\affiliation{John Adams Institute for Accelerator Science and Department of Physics,University of Oxford, Denys Wilkinson Building, Keble Road, Oxford OX1 3RH, United Kingdom}%
\author{R. Walczak}%
\email{roman.walczak@physics.ox.ac.uk}
\affiliation{John Adams Institute for Accelerator Science and Department of Physics,University of Oxford, Denys Wilkinson Building, Keble Road, Oxford OX1 3RH, United Kingdom}%

\date{\today}

\begin{abstract}
We describe a new approach for driving GeV-scale plasma accelerators with long laser pulses. We show that the temporal phase of a long, high-energy driving laser pulse can be modulated periodically by co-propagating it with a low-amplitude plasma wave driven by a short, low-energy seed pulse. Compression of the modulated driver by a dispersive optic generates a train of short pulses suitable for resonantly driving a plasma accelerator. Modulation of the driver occurs via well-controlled linear processes, as confirmed by good agreement between particle-in-cell (PIC) simulations and an analytic model. PIC simulations demonstrate that a \SI{1.7}{J}, \SI{1}{ps} driver and a \SI{140}{mJ}, \SI{40}{fs} seed pulse can accelerate electrons to energies of \SI{0.65}{GeV} in a plasma channel with an axial density of $\SI{2.5E17}{cm^{-3}}$. This work opens a route to high-repetition-rate, GeV-scale plasma accelerators driven by thin-disk lasers, which can provide joule-scale, picosecond-duration laser pulses at multi-kilohertz repetition rates and high wall-plug efficiencies.

This article was published in Physical Review Letters (Vol. 127, No. 18) on 26$^{th}$ of October 2021. DOI: 10.1103/PhysRevLett.127.184801 \copyright 2021 American Physical Society. The journal version should be used for citations. 
\end{abstract}

\maketitle

The acceleration fields generated in a laser-plasma accelerator (LPA) are of the order of the wave-breaking field $E_{wb} = m_e c \omega_p / e$, and for parameters of interest \cite{Tajima1979} can reach $\SI{100}{GV.m^{-1}}$  --- around three orders of magnitude greater than possible in  radio-frequency accelerators. The duration of the particle bunches accelerated by a LPA are a fraction of the plasma period $T_p = 2 \pi / \omega_p$, i.e.\ a few femtoseconds \cite{Faure2006, Tilborg:2006,Ohkubo:2007, Debus:2010, Lundh2011, Buck2013, Heigoldt:2015cd}. Here, $\omega_p = (n_e e^2 / m_e \epsilon_0)^{1/2}$ is the plasma frequency, where $n_e$ is the electron density. 

The development of LPAs has seen remarkable progress in recent years \cite{Esarey2009, Hooker2013a, Joshi2020}. Micrometer-sized electron bunches with GeV-scale energies can be generated from LPA stages a few centimetres long  \cite{Leemans:2006, Kneip:2009, Wang:2013el, Leemans:2014kp, Gonsalves:2019ht}, and hence LPAs have the potential to drive highly compact sources of high-energy, femtosecond-duration particles and radiation \cite{Kneip:2010, Schlenvoigt:2008, Fuchs:2009, Phuoc:2012vb, Corde2013,Khrennikov:2015gx, Wang2021}. Indeed, their proof-of-principle applications to X-ray imaging \cite{Kneip2011,Hussein2019,Ben-Ismail2011}, photochemistry \cite{Gauduel2010}, and electron radiotherapy \cite{Labate2020,Polanek2021,Svendsen2021} have been demonstrated, and conceptual designs for LPA facilities have been developed \cite{Alegro,EuPRAXIA2020}.

Efficient excitation of the plasma wave requires that the duration of the driving laser pulse satisfies $\tau < T_p/2$, and the peak laser intensity is of order $\SI{E18}{W.cm^{-2}}$. Few laser systems can meet these challenging requirements, and in practice almost all LPAs are driven by Ti:sapphire laser pulses with $\tau \approx \SI{40}{fs}$ and peak powers up to the PW range. These lasers are limited \cite{Dawson2012} to pulse repetition rates $f_\mathrm{rep}\lesssim \SI{10}{Hz}$ and mean powers $\lesssim \SI{100}{W}$, whereas many applications of LPAs would require operation at $f_\mathrm{rep} \gtrsim \SI{1}{kHz}$, with mean laser powers in the kW range. To date, kHz operation of LPAs \cite{He2013,Guenot2017,Salehi2017,Gustas2018,Rovige2020,Salehi2021} has been limited to generating electron energies below \SI{15}{MeV} by the low pulse energies available from kHz Ti:sapphire lasers.

 \begin{figure*}[tb]
     \centering
     \includegraphics[width=\textwidth]{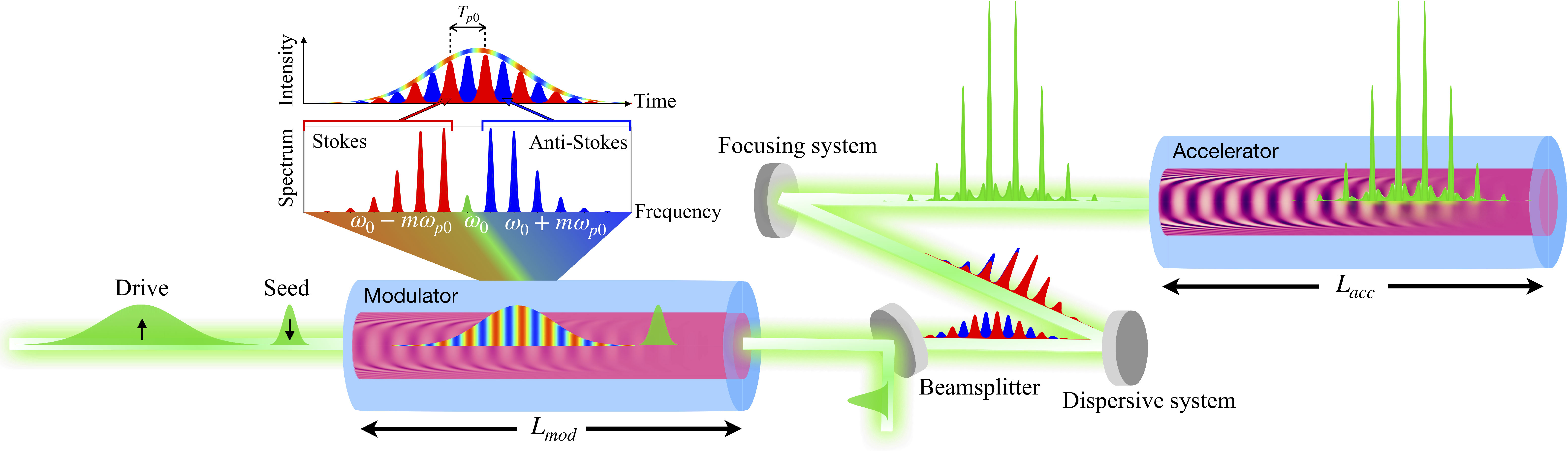}
    \caption{Schematic diagram of a LPA driven by plasma-modulated laser pulses. A long, high-energy, drive laser pulse is phase-modulated in the modulator stage by its interaction with the plasma wave driven by a short, low-energy seed pulse. The modulation generates sidebands at $\omega_0 + m \omega_\mathrm{p0}$, although the temporal intensity profile of the drive pulse remains smooth. After leaving the modulator stage, the seed pulse is removed by a polarizing beam-splitter, and the drive pulse is passed through (or reflected from) a dispersive optical system which removes the relative spectral phase of the sidebands, to form a train of short pulses spaced by $T_\mathrm{p0} = 2 \pi / \omega_\mathrm{p0}$. This pulse train is focused into an accelerator stage, which comprises a plasma channel with the same on-axis density as that of the modulator stage. The pulse train resonantly excites a strong plasma wave that can be used for particle acceleration.} 
     \label{Fig:Concept}
 \end{figure*}

A further consideration of vital importance for future high-mean-power LPAs is the wall-plug efficiency. For Ti:sapphire lasers this is $<0.1\%$ \cite{UkRoadmap}, and with laser-to-wake and wake-to-bunch efficiencies each less than 50\% \cite{Streeter2020}, the overall efficiency is far from the $>10\%$ required by future, cost-competitive high-energy LPAs \cite{Albert2021}.

Substantially more efficient lasers have been developed in recent years. For example, thin-disk lasers have optical-to-optical efficiencies exceeding 50\% and have recently generated pulse energies of $\sim \SI{1}{J}$ at $f_\mathrm{rep} = \SI{1}{kHz}$ \cite{Herkommer:2020, Produit2020, Wang:2020}. However, their long (picosecond) pulse duration makes them unsuitable for driving LPAs directly. Spectral broadening in gases has been shown \cite{Rocca2021,Fan2021, Kaumanns2021} to reduce the duration of thin-disk laser pulses, but sub-\SI{100}{fs} pulses with the energy required to drive high-energy LPAs have yet to be demonstrated.

Many approaches for driving LPAs with long ($\tau > T_p$) pulses have been investigated. In the plasma beat-wave accelerator (PBWA) \cite{Kitagawa1992} two long pulses of angular frequencies $\omega_1$ and $\omega_2 = \omega_1 + \omega_p$ are combined to create a modulated pulse which can resonantly excite a plasma wave. However, current high-efficiency laser technologies cannot generate a suitable second wavelength with the required high average power. Plasma waves can also be driven via self-modulation of long laser pulses \cite{Andreev92,King2021}, but this relies on stochastic, and highly nonlinear, processes which would make controlled injection and acceleration difficult.

In this Letter we describe a new approach which could be used to drive GeV-scale, multi-kHz LPAs with single, joule-level, picosecond-duration pulses of the type recently demonstrated by thin-disk lasers. We use a one-dimensional (1D) analytic model and particle-in-cell (PIC) simulations to demonstrate the physics underlying this scheme. To provide an example, we simulate the acceleration of externally-injected electrons by a \SI{1.7}{J}, \SI{1}{ps} drive laser pulse to a mean energy of \SI{0.65}{GeV}.

As shown in Fig.\ \ref{Fig:Concept}, our approach has three stages. In the first, the modulator stage, a short ($\tau_\mathrm{seed} \lesssim T_{p0}/2$), low energy seed pulse and a long ($\tau_\mathrm{drive} \gg T_{p0}$), high-energy driving laser pulse are focused into a plasma waveguide. Here $T_{p0}$ refers to the plasma period on-axis.  The seed pulse drives a low amplitude plasma wave which periodically modulates the temporal phase of the drive pulse, and hence generates frequency sidebands at $\omega_0 + m \omega_{p0}$. If isolated, the red- ($m <0$) and blue-shifted ($m > 0$) sidebands would form a pair of \emph{temporal} pulse trains separated by $T_{p0} / 2$, but with both sets present the temporal profile of the drive pulse remains smooth. 

In the second stage, after removal of the seed pulse, the spectrally-modulated drive pulse is converted into a temporally-modulated train of short pulses, separated by $T_{p0}$, by introducing a shift of $T_{p0} / 2$ between the red- and blue-shifted trains. This can be achieved by introducing a dispersive optical system \cite{Hadrich2016,Grebing2020,Alessi2019} with the correct group delay dispersion (GDD). We note that a similar technique was proposed for plasma-based pulse compression in PBWA through electromagnetic (EM) cascading \cite{Kalmykov2005,Kalmykov2006}. A major advantage of our scheme is that only a single high-energy drive pulse is required, rather than two pulses with a frequency difference matched to $\sim\omega_{p0}$.

In the third (accelerator) stage, the pulse train is focused into a second plasma waveguide with the same axial density as the modulator, which resonantly excites a large-amplitude plasma wave for acceleration of externally injected electrons.

Considerable insight into these processes can be gained with the aid of a 1D analytic model (see Supplemental Material \cite{supp_mat}). The drive pulse has a normalized vector potential, $a(z,t) = b(z,t) \exp\left[ \mathrm{i} (k_0 z - \omega_0t)\right]$, where $\omega_0$ is the centre frequency of the pulse, and $z$ is the coordinate along the propagation axis. After co-propagating a distance $z$ with  a linear plasma wave of density $n_e(z,t) = n_0 + \delta n \cos \left( k_{p 0} z - \omega_{p 0} t + \Delta \phi \right)$, the envelope of the pulse becomes \cite{Esarey:1990ve},

\begin{align}
    b(\zeta, \tau) \approx \left| b(\zeta,0)\right|  \sum_{m = -\infty}^\infty  \mathrm{i}^m J_m(- \beta)  \exp \left[\mathrm{i} m \left( \omega_{p 0} \tau + \Delta \phi ' \right) \right].
\end{align}
Here $\zeta = z - v_g t$, where $v_g$ is the group velocity of the driver, $\tau = t$, $\beta = (1/2) (\omega_{p 0 }^2 / \omega_0)(\delta n / n_0) (z / v_g)$, and $\Delta \phi' = -(\Delta \phi + \omega_{p 0} z/ v_p)$, where $v_p$ is the phase velocity of the plasma wave. The generation of sidebands at $\omega_0 + m \omega_{p 0}$ is immediately apparent. In the temporal domain, the red- and blue-shifted sidebands each form trains of pulses separated by $T_{p0}$, with a relative shift \cite{supp_mat} between the two trains of $T_{p0}/2$.

\begin{figure*}[tb]
     \centering
     \includegraphics[width=\textwidth]{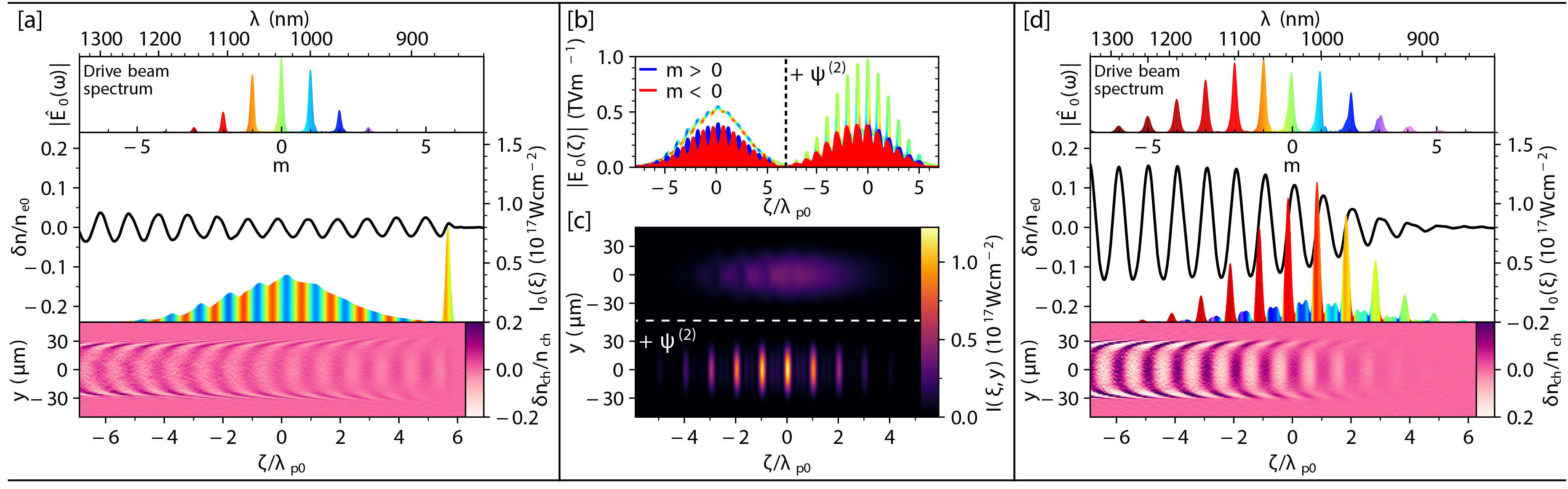}
     \caption{Particle-in-cell simulations. (a) PIC simulations of the modulator stage, with results shown for the end of the modulator ($z=\SI{120}{mm}$). The top panel shows the on-axis spectral intensity of the drive pulse, plotted against a frequency scale $m = (\omega - \omega_0) / \omega_\mathrm{p0}$. The middle panel shows the longitudinal intensity profiles of the seed and driver pulses and the relative amplitude of the plasma wave on axis ($y = 0$).  The bottom panel shows, for $z = \SI{120}{mm}$, a 2D plot of the electron density relative to the channel profile $\delta n_{ch} / n_{ch}=(n_e-n_{ch})/n_{ch}$. The shading of the longitudinal profiles indicates the local effective frequency $- \mathrm{d} \phi / \mathrm{d} t$, where $\phi$ is the temporal phase, using the same color scale as the top panel. 
     (b) The modulus of the on-axis electric field of the drive pulse together with the red- and blue-shifted components before and after application of a quadratic spectral phase $\psi^{(2)} = \SI{- 1480}{fs^2}$. (c) The corresponding 2D intensity profiles. 
     (d) The same plots as in (a) but at a distance $z = \SI{50}{mm}$ into the accelerator stage.}
     \label{fig:StartToEnd}
 \end{figure*}

To demonstrate this scheme, and to gain further insight, we performed high-resolution 2D PIC simulations with the EPOCH code \cite{Arber:2015hc}. We compared these results with those obtained with a different PIC code implemented with cylindrical symmetry to check the validity of the 2D simulations   \cite{supp_mat}. The parameters of the drive and seed laser pulses were chosen to be similar to those recently demonstrated for thin-disk lasers operating at $f_\mathrm{rep} \gtrsim \SI{1}{kHz}$ at a wavelength of $\lambda_0=\SI{1030}{nm}$. Both pulses were assumed to be bi-Gaussian, with a drive (seed) energy of $E^{\text{drive}}_p=\SI{600}{mJ}$ ($E^{\text{seed}}_p=\SI{50}{mJ}$) and a full-width at half maximum pulse duration of $\tau_\mathrm{drive} = \SI{1}{ps}$ ($\tau_\mathrm{seed} = \SI{40}{fs}$). The pulses were separated by \SI{1.7}{ps}. 

The plasma parameters of the modulator and accelerator stages were identical. For radii $r < r_0 = 1.2w_M$, the  electron density was $n_{ch}(r) = n_{e0} + (1/\pi r_e  w_M^2)  (r / w_M)^{\alpha}$, where $n_{e0} \equiv n_{e}(0) = \SI{2.5E17}{cm^{-3}}$, and $w_M = \SI{30}{\mu m}$ is approximately the spot size of the lowest-order channel mode.  For $r > 1.2w_M$, $n_e(r) = n_e(r_0)$ for $\Delta r = \SI{10}{\mu m}$, before decreasing linearly to zero in the same distance. The parameter $\alpha = 10$, and hence the channels were steep-sided, with finite losses.

The drive and seed pulses were focused to a spot-size $w_0 = w_M = \SI{30}{\mu m}$ at the entrance to the modulator channel. Figure\ \ref{fig:StartToEnd} (a) shows that the seed pulse drives a plasma wave of amplitude $\delta n/n_{e0} \approx 2\%$ on axis, which generates side-bands on the drive pulse at $\omega_0 + m \omega_\mathrm{p 0}$, although its temporal profile remains smooth. Figures \ref{fig:StartToEnd}(b,c) show, after the modulator, the on-axis electric field (b) and temporal intensity profile (c) of the drive pulse, before and after the application of a quadratic spectral phase with a group delay dispersion (GDD) of $\psi^{(2)} = -\SI{1480}{fs^2}$. This value of the GDD was determined numerically to produce the highest-intensity pulse train, and is in excellent agreement with the value  $\psi^{(2)}_\mathrm{opt} = \pm \SI{1485}{fs^2}$ predicted by the 1D  model \cite{supp_mat}. It is evident that the dispersion converts the phase-modulated, but temporally-smooth, drive pulse into a train of short pulses separated by $T_{p0}=\SI{220}{fs}$. The use of seed and drive pulses of different wavelengths is explored in \cite{supp_mat}.

Figure \ref{fig:StartToEnd}(d) shows the drive pulse train, and the plasma wake, after \SI{50}{mm} propagation in the accelerator stage. The pulse train resonantly excites a strong plasma wave with $\delta n/n_e \approx 15\%$ over the whole length of the accelerator. The spectrum of the driver is further modulated by this plasma wave, leading to the formation of additional sidebands through EM cascading \cite{Kalmykov2006}; it is also red-shifted, the shift increasing towards the back of the pulse train, where the wake amplitude is higher.

\begin{figure}[tb]
     \centering
     \includegraphics[width=\columnwidth]{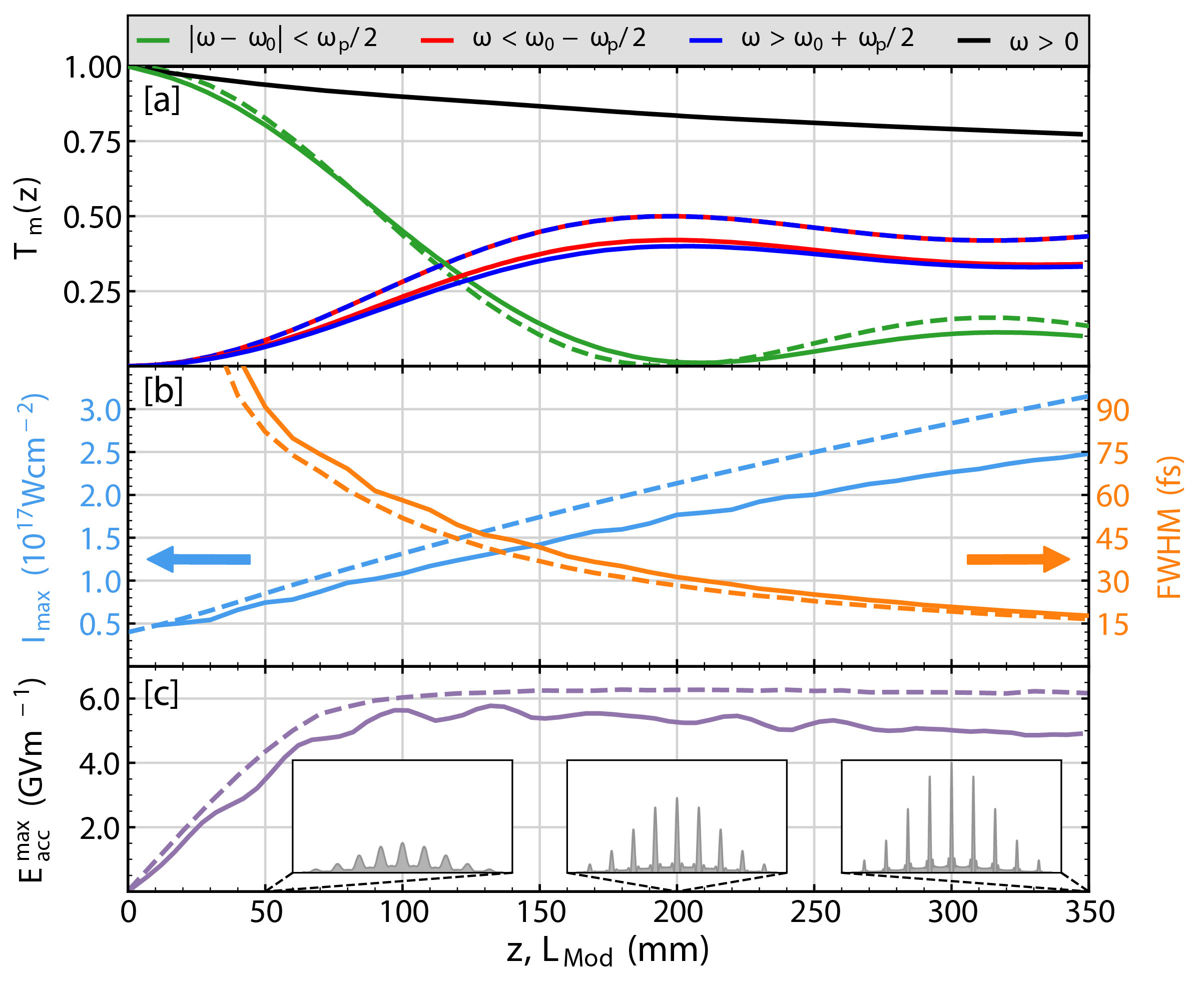}
     \caption{[Color online]  Comparison of the results of the 1D analytic model (dashed lines) and PIC simulations (solid lines), plotted as  a function of the length of the modulator stage. (a) The relative transmitted energies of the drive pulse (black), and of its components  in the central band (i.e.\ $|\omega - \omega_0| < \omega_p /2$, green), and in the blue-shifted (i.e.\ $|\omega - \omega_0| > \omega_p /2$, blue) and red-shifted ($|\omega - \omega_0| < -\omega_p /2$, red) sidebands. (b) The peak intensity (blue), and FWHM duration (orange), of the most intense pulse in the train generated by applying a quadratic spectral phase, optimized to yield the highest-intensity pulse train, on the drive pulse emerging from the modulator stage. (c) The peak accelerating electric field produced by injecting into the accelerator stage the pulse trains which would be generated by compressing the drive pulse at that point in the modulator.}
     \label{fig:PhotonAcc}
 \end{figure}

 The results of the PIC simulation can be compared with the 1D model of the modulator stage \cite{supp_mat}. Figure \ref{fig:PhotonAcc}(a) shows, as a function of the length $L_\mathrm{mod}$ of the modulator, the relative energies in the central, red-shifted, and blue-shifted bands of the drive pulse. It can be seen that the analytic model and PIC simulations are in close agreement. The main difference is that leakage in the plasma channel, which is not included in the 1D model,  attenuates the drive pulse in the PIC simulation. Figures \ref{fig:PhotonAcc}(b, c) show: (b) the properties of the pulse train which would be produced if the drive pulse at that point was compressed by introducing a GDD optimized to yield the highest-intensity pulse train; and (c) the peak accelerating field produced by injecting these trains into the accelerator stage. Examples of the pulse trains generated for three values of $L_\mathrm{mod}$ are shown in the insets of Fig.\ \ref{fig:PhotonAcc}(c). Again, very good agreement is obtained. The peak pulse intensity is seen to grow approximately linearly with $L_\mathrm{mod}$; this is expected from the 1D model, since the spectral bandwidth increases as $L_\mathrm{mod}$, and hence the duration of each compressed pulse will vary as $L_\mathrm{mod}^{-1}$. It can be seen that trains of pulses as short as \SI{15}{fs} can be generated.  Although the peak intensities of the pulses increase linearly with $L_\mathrm{mod}$, the peak accelerating field does not increase significantly for $L_\mathrm{mod} \gtrsim \SI{100}{mm}$. This is expected since for single, short ($\tau_\mathrm{drive} \ll T_\mathrm{p0}$) drive pulses the wake amplitude depends only on the energy of the pulse \cite{Dorchies1999}, and for resonant pulse trains the wake amplitude depends only on the total energy of the train. The excellent agreement between the 1D analytic model and the PIC simulations demonstrates that, at least for this parameter range, the former captures the key physics involved in the operation of the modulator stage.

The analytic model and PIC simulation presented above demonstrate that plasma modulation can be used to convert a high-energy, long drive pulse to a train of short pulses suitable for resonant excitation of a large amplitude plasma wave. We emphasize that processes by which the drive pulse is modulated are linear, and can therefore be well controlled. Nevertheless it is important to avoid instabilities developing in either the modulator or accelerator stages. In this work the growth of instabilities, such as forward Raman scattering \cite{Decker1996, Mori1997}, was suppressed by operating in a favourable parameter regime, and by employing a plasma channel with small, but finite losses. These losses damp higher-order waveguide modes excited by the interaction between the drive pulse and the seed-driven plasma wave,  preventing modulation of the temporal profile of the drive pulse, which would be susceptible to instability-driven growth \cite{Antonsen1995}.

\begin{figure}[tb]
     \centering
     \includegraphics[width=\columnwidth]{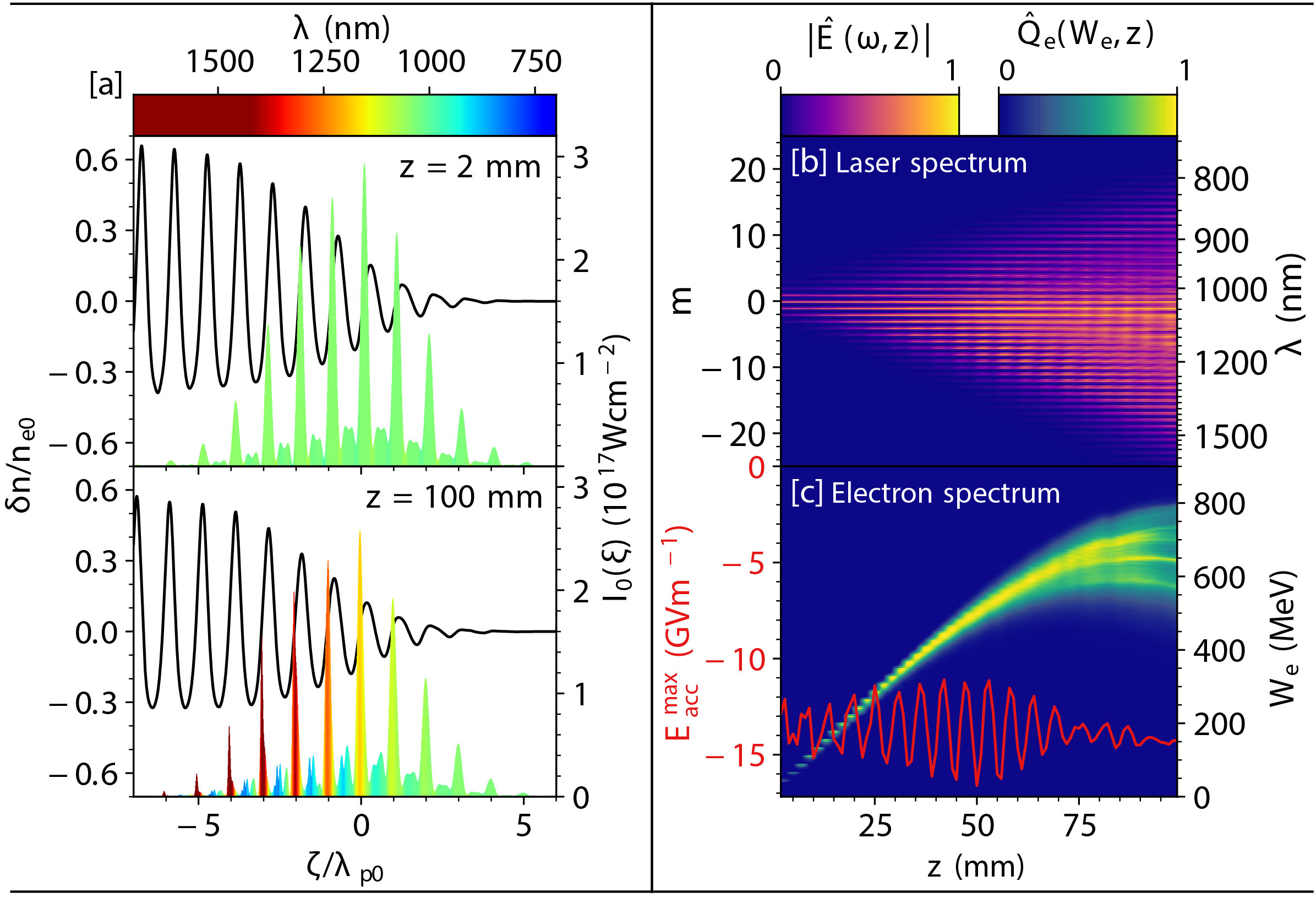}
     \caption{[Color online] Performance of a scaled accelerator with seed and drive pulse energies increased to \SI{140}{mJ} and \SI{1.7}{J} respectively. (a) The on-axis longitudinal profiles of the laser intensity and the relative electron density $\delta n/n_0$ at $z =  \SI{2}{mm}$ (top) and $z = \SI{100}{mm}$ (bottom) in the acceleration stage. The color scale shows the local laser wavelength. (b) Evolution of the normalized spectral intensity of the drive laser with propagation distance $z$ in the acceleration stage. (c) Evolution of the normalized energy spectrum $ \hat{Q}_e (W_e , z)$ of the injected electron bunch with $z$.}
     \label{fig:AccelerationStage}
 \end{figure}

In Fig.\ \ref{fig:AccelerationStage} we demonstrate that the scheme can be scaled to higher drive pulse energies, and hence higher acceleration gradients, whilst ensuring that the modulator operates in a well-controlled, linear regime. For these simulations the parameters of the accelerator stage, and the axial density of the modulator stage, were the same as in Fig.\ \ref{fig:StartToEnd}. However, the matched spot size of the modulator was increased to $\SI{50}{\micro m}$ in order to allow the seed and drive pulse energies to be increased to \SI{140}{mJ} and \SI{1.7}{J} whilst keeping their peak intensities the same as in Fig.\ \ref{fig:StartToEnd}. Figure \ref{fig:AccelerationStage}(a) shows clearly that a quasi-linear wakefield, with an amplitude considerably larger than that of Fig.\ \ref{fig:StartToEnd}, can be driven over the entire length of the 100mm acceleration stage.

To demonstrate particle acceleration, a \SI{1}{pC} electron bunch of energy 35 MeV, \SI{5}{fs} root-mean-square duration, and $\SI{4}{\mu m}$ transverse width was injected into the focusing phase of the wakefield, at the position of peak acceleration. Figure \ref{fig:AccelerationStage}(c) shows the evolution with $z$ of the normalized energy spectrum $\hat{Q}_e (W_e,z)$ of this bunch, where $W_e$ is the electron energy. It can be seen that the bunch maintains a relatively narrow energy spectrum up to  $z \approx \SI{50}{mm}$, at which point the mean energy is $\sim\SI{500}{MeV}$. At larger $z$, dephasing causes the energy spectrum to broaden. The mean electron energy at the end of the accelerator stage is \SI{0.65}{GeV}. The laser-plasma energy transfer in the accelerator stage is found \cite{supp_mat} to be $9\%$.

The results of Fig.\ \ref{fig:AccelerationStage}  show clearly that the concept described in this paper can be scaled to generate higher particle energies without introducing unwanted instabilities in either modulator or acceleration stages. Our results serve to demonstrate the operation of this scheme, but further work will be required to fully explore its potential. For example, it is likely that the energy transfer efficiency could be increased, and the properties of the accelerated bunch improved, by optimizing the parameters of the drive laser and plasma channel, and the beam loading of the plasma wakefield.

In summary, we have presented a new approach for resonantly driving plasma accelerators with long laser pulses, based on co-propagation with a low-amplitude plasma wave and compression by a dispersive optic of the spectrally-modulated output to form a train of short pulses. This scheme was demonstrated via a 1D analytic model and 2D PIC simulations, which were found to be in excellent agreement. Numerical simulations showed that a quasilinear wakefield could be driven over a dephasing length to accelerate a test bunch to a mean energy $\sim \SI{0.65}{GeV}$.

We note that the key components required to realize this new approach have all recently been demonstrated, and in principle all are capable of multi-kilohertz operation. These include: metre-scale, low-loss, all-optical plasma waveguides \cite{Shalloo:2018fy, Shalloo:2019hv, Smartsev2019, Picksley2020_1, Picksley:2020, Feder2020, Miao:2020}; \SI{100}{mJ}-scale, sub-\SI{50}{fs} seed laser pulses \cite{ThalesLaserGroup}; and joule-level, few-picosecond, \SI{1030}{nm} drive laser pulses \cite{Herkommer:2020, Wang:2020}. The proposed scheme requires tight temporal and spatial overlap of multiple laser beams, the most challenging of which are: (i) control of the delay between the electron bunch and the seed pulse ($\sim \SI{10}{fs}$); and (ii) control of the pointing of the seed and drive pulses relative to the waveguide axes ($<\SI{10}{\micro rad}$). These requirements have already been met \cite{Genoud2011,Picksley2020_1,Picksley:2020,Miao:2020,Maier2020,Jalas2021,Wu2021} in experiments operating at $f_\mathrm{rep} = \SI{1}{Hz}$, and even tighter tolerances could be achieved with the improved feedback made possible by kHz operation.

\vspace{10mm}
\textbf{Acknowledgements}
The authors acknowledge useful discussions with A. Picksley and J.A. Holloway. This work was supported by the UK Science and Technology Facilities Council (STFC UK) [grant numbers ST/P002048/1 and ST/V001655/1]; the Engineering and Physical Sciences Research Council (EPSRC UK) [grant numbers EP/V006797/1 and EP/R513295/1]. This material is based upon work supported by the Air Force Office of Scientific Research under award number FA9550-18-1-7005. 
This work was supported by the European Union's Horizon 2020 research and innovation programme under grant agreement No. 653782. This work required significant computing resources which were funded by the plasma HEC Consortium [EPSRC grant number EP/R029149/1] and UKRI funding [ARCHER2 Pioneer Projects]. The development of the EPOCH code is supported in  part by the UK EPSRC grants  EP/G054950/1,EP/G056803/1, EP/G055165/1 and EP/ M022463/1. Computing resources were provided by ARCHER and ARCHER2 [ARCHER2 PR17125] UK supercomputers as well as STFC Scientific Computing Department’s SCARF cluster.

\bibliography{PlasmaModulatedPulses.bib}

\newpage
\setcounter{page}{0} 

\begin{center}
{\bfseries GeV-scale accelerators driven by plasma-modulated pulses from kilohertz lasers --- Supplemental Material}
\vspace{14pt}

{O. Jakobsson, S.M. Hooker and R. Walczak}%

\vspace{10pt}

\textit{John Adams Institute for Accelerator Science and Department of Physics,University of Oxford, Denys Wilkinson Building, Keble Road, Oxford OX1 3RH, United Kingdom}%
\end{center}

In this document we provide additional information on the analytic theory and Particle-In-Cell (PIC) simulations presented in the main paper.

\section{Analytic theory of the modulator}
Here we provide further information on the one-dimensional (1D) linear theory of the modulator. We follow the approach of Esarey et al.\ \cite{Esarey:1990ve} and describe the drive pulse by the normalized vector potential,

\begin{align}
	a(z,t) = b(z,t) \exp\left[ \mathrm{i} (k_0 z - \omega_0t)\right],
\end{align}
    where $z$ is the coordinate along the axis of propagation. Here, $\omega_0$ and $k_0$ are the central angular frequency and wave number of the pulse in the absence of a plasma wave, which satisfy $\omega_0^2 = c^2 k_0^2 + c^2 k_{p0}^2$, in which  $k_{p0}^2 = \omega_{p0}^2 / v_p^2$, and where $\omega_{p0}^2 = n_0 e^2 / m_e \epsilon_0$,  $n_0$ being the ambient electron density and $v_p$ is the phase velocity of the plasma wave.
    
    We make the slowly-varying envelope approximation, and transform to the coordinates $(\zeta, \tau)$, where the local coordinate $\zeta = z - v_g t$, in which $v_g$ is the group velocity of the radiation pulse, and $\tau = t$. We then find, \cite{Esarey:1990ve}

\begin{align}
    b(\zeta, \tau) = \left| b_0(\zeta)\right|	\exp \left[ - \mathrm{i} \frac{c^2}{2 \omega_0} \int_0^\tau \mathrm{d} \tau' \delta k_p^2(\zeta, \tau')\right],\label{Eqn:Gen_soln}
\end{align}
in which $b_0(\zeta) = b(\zeta, 0)$, and,

\begin{align}
	\delta k_p^2(\zeta, \tau) &= k_p^2 - k_{p 0}^2 \approx k_{p 0}^2 \left[\frac{n_e(\zeta, \tau)}{n_0} - 1 \right],\label{Eqn:kp-squared}
\end{align} 
where $n_e(\zeta, \tau)$ is the local  electron density in the presence of the plasma wave. The approximation ignores relativistic effects.

We wish to consider the effect on the driver of co-propagation with a plasma wave described by,

\begin{align}
	n_e(z,t) &= n_0 + \delta n \cos \left( k_{p 0} z - \omega_{p 0} t + \Delta \phi \right) \nonumber\\
	&= n_0 + \delta n \cos \left[ \frac{\omega_{p 0}}{v_p}\left(\zeta +\Delta v \tau \right) + \Delta \phi \right],
\end{align}
where $\Delta v =v_g - v_p $ and $\Delta \phi$ is a constant phase shift. Substituting this into eqns \eqref{Eqn:Gen_soln} and \eqref{Eqn:kp-squared} yields,

\begin{align}
b(\zeta, \tau) &= \left| b_0(\zeta)\right|	\exp \left\{ - \mathrm{i} \frac{1}{2} \frac{\omega_{p 0 }^2}{\omega_0}\frac{\delta n}{n_0} \tau \sinc \left(\frac{1}{2} \frac{\omega_{p 0}}{v_p} \Delta v \tau \right)\right. \nonumber\\
 &\times \left. \cos \left[ \frac{\omega_{p 0}}{v_p}\zeta   + \frac{1}{2}\frac{\omega_{p 0}}{v_p} \Delta v \tau + \Delta \phi \right] \right\},\label{Eqn:FullExpression}
\end{align}
where $\sinc x =  (\sin x )/ x$. We note that the magnitude of the phase modulation \emph{grows linearly with $\tau$} when $\Delta v = 0$. 

If the difference between the group velocity of the driver and the phase velocity of the plasma wave is small, i.e.\ if $\Delta v\tau \ll \lambda_p$, where $\lambda_{p0}=2\pi/k_{p0}$, then,

\begin{align}
b(\zeta, \tau) &\approx \left| b_0(\zeta)\right| \exp \left\{ - \mathrm{i} \frac{1}{2} \frac{\omega_{p 0 }^2}{\omega_0}\frac{\delta n}{n_0} \tau \cos \left[ \frac{\omega_{p 0}}{v_p}\zeta   + \Delta \phi \right] \right\}.
    \label{Eqn:SpectralShiftSameFrequency}
\end{align}
For the times of interest, i.e.\ times corresponding to the arrival of the pulse, we have $\tau \approx z/v_g$. Hence, making use of the Jacobi-Anger expansion, we can write,

\begin{align}
b(\zeta, \tau) &\approx \left| b_0(\zeta)\right| \exp \left[ - \mathrm{i} \frac{1}{2} \frac{\omega_{p 0 }^2}{\omega_0}\frac{\delta n}{n_0} \frac{z}{v_g} \cos \left(\omega_{p 0} \tau + \Delta \phi '\right) \right] \nonumber\\
&\approx \left| b_0(\zeta)\right|  \sum_{m = -\infty}^\infty \mathrm{i}^m J_m(- \beta)\exp \left[\mathrm{i} m \left( \omega_{p 0} \tau + \Delta \phi ' \right) \right],\label{Eqn:b_out_with_sidebands}
\end{align}
where $\Delta \phi' = -(\Delta \phi + \omega_{p 0} z/ v_p)$ and 

\begin{align}
	\beta = \frac{1}{2} \frac{\omega_{p 0 }^2}{\omega_0}\frac{\delta n}{n_0} \frac{z}{v_g}.
\end{align}
It is immediately clear that co-propagation of the driver with the plasma wave causes the driver to develop sidebands at frequencies given by $\omega_0 + m \omega_{p 0}$.

If we make use of the relations,

\begin{align}
	J_{-m}(x) &= (-1)^m J_m(x)\label{Eqn:Bessel_identity_1}\\
	J_m(-x) &= -J_{-m}(x),\label{Eqn:Bessel_identity_2}
\end{align}
then the sum in the above can be re-written as:

\begin{align}
	&\sum_{m = -\infty}^\infty i^m J_m(- \beta)\exp \left[i m \left( \omega_{p 0} t + \Delta \phi ' \right) \right] \nonumber\\
	&= - \sum_{m = -\infty}^\infty J_{m}(\beta)\exp \left[-i m \left( \omega_{p 0} t - k_{p 0} z - \Delta \phi + \frac{\pi}{2}\right) \right],
\end{align}
where we have substituted for $\Delta \phi'$. We see that the sum represents a set of blue- ($m >0$) and red-($m<0$) sidebands.

\subsection{Sideband phase and required GDD}
The spectral phase of the blue-shifted sidebands is easily seen to be,

\begin{align}
    \psi_m^+ = \theta_m + m \left(k_{p0} z - \frac{\pi}{2} + \Delta \phi \right)
\end{align}
where we define $\theta_m$ as the argument of $J_{|m|}(\beta)$. Taking into account \eqref{Eqn:Bessel_identity_1}, the phase of the red-shifted sidebands can be shown to be 

\begin{align}
    \psi_m^- = \theta_m + m \left(k_{p0} z - \frac{\pi}{2} + \Delta \phi \pm \pi\right),
\end{align}
so that, in general, the phase of sideband $m$ is,

\begin{align}
    \psi_m &= \theta_m + m \left(k_{p0} z + \Delta \phi \right) - |m| \frac{\pi}{2}\\
    &\approx  m \left(k_{p0} z + \Delta \phi \right) - |m| \frac{\pi}{2},
\end{align}
where the approximation holds for $\beta < x_{00}$, where $x_{00} \approx 3.83171$ is the first non-zero solution of $J_1(x_{00})=0$, since then $\theta_m = 0$ for all $m$.

From the Fourier shift theorem, the term linear in $m$ (and hence in frequency) corresponds to a shift in the arrival time with $z$. However, the term $- |m| \pi / 2$ is \emph{not} linear in frequency since red- and blue-shifted sidebands of the same order $|m|$ have the same phase.

Optimum compression of the spectrally-modulated driver would be obtained if this nonlinear phase could be removed by the dispersive optical system, i.e.\ if the system could introduce a phase shift of $+ |m| \pi / 2$. In practice,  dispersive systems typically introduce a spectral phase of the form $\psi(\omega) = \psi^{(0)} + \psi^{(1)}(\omega - \omega_0) + (1/2) \psi^{(2)}(\omega - \omega_0)^2 \ldots $, where the $\psi^{(n)}$, $n = 0,1,2, \ldots$ are coefficients.

The linear term is unimportant since it only affects the arrival time of the pulse train. We therefore consider the next term, the group delay dispersion (GDD), $\psi^{(2)}$, and optimize its value. We do this by constructing the function,

\begin{align}
	\chi^2 &= \sum_{m = -N}^{+N} |J_m(\beta)|\left[\frac{1}{2}\psi^{(2)}(m \omega_{p 0} )^2 - (-\psi_m)\right]^2,
\end{align}
i.e. the mean square difference between the phase shift introduced by the dispersive system, and that desired, weighted by the strength of the side-band. Minimizing $\chi^2$ with respect to $\psi^{(2)}$ we find, for $\beta \lesssim 3.83171$,

\begin{align}
\psi^{(2)}_\mathrm{opt}  = \pm \frac{\sum_{m = 1}^{N}|J_m(\beta)|m^3}{\sum_{m = 1}^{N} |J_m(\beta)|m^4 } \frac{\pi}{\omega_{p 0}^2}.
\end{align}
The positive (negative) GDD compensates for the negative (positive) spectral chirp about the peaks (troughs) of the plasma wake.

\subsection{Pulse arrival time}
Using the results above we can write \eqref{Eqn:b_out_with_sidebands} as,

\begin{align}
b(\zeta, \tau) &\approx 
 - \left| b_0(\zeta)\right| \nonumber\\
 &\times \sum_{m = -\infty}^\infty  |J_m(\beta)|\exp \left[\mathrm{i} m \left(k_{p0}z - \omega_{p0} t + \Delta \phi \mp  \frac{\pi}{2}  \right)\right],
\end{align}
where the upper and lower signs hold for $m>0$ and $m<0$ respectively. The sum will be a maximum when the term in curved brackets is equal to $2 \pi$, from which we deduce that peaks will occur when,

\begin{align}
t_n &=  n T_{p 0} + \frac{z}{v_p} + \frac{\Delta \phi}{\omega_{p0}} \mp \frac{T_{p 0}}{4},
\end{align}
where $T_{p 0} = 2 \pi / \omega_{p0}$ and $n$ is an integer. Hence we see that the red- and blue-shifted sidebands each correspond to a train of  pulses, separated in time by $T_{p 0} / 2$.

\subsection{Wake estimate}
The wakefield driven by the seed pulse, and compressed pulse trains, is estimated using one-dimensional quasistatic fluid theory \cite{Esarey:2009}. A laser with a pulse profile described by the normalized vector potential $a = eA/m_e c^2$ will excite a wakefield with a wake potential $\phi$ described by
\begin{equation}
    \frac{\partial^{2} \phi}{\partial t^{2}}= \frac{\omega_{p0}^{2}}{2}\left[\frac{1+a^{2} / 2}{(1+\phi)^{2}}-1\right],
    \label{Eqn:WakePhi}
\end{equation}
where the longitudinal electric wakefield $E_{z}=-\frac{m_{e} c^{2}}{e} \frac{\mathrm{d} \phi}{\mathrm{d} z}$. The corresponding plasma-density perturbation is,

\begin{equation}
    \frac{\delta n}{n_{e0}} = \frac{1+a^{2} / 2+(1+\phi)^{2}}{2(1+\phi)^{2}}-1.
    \label{Eqn:WakeDeltan}
\end{equation}
The spectral modulation of the long pulse in the modulator stage was calculated by first estimating the wake amplitude $\delta n/n_{e0}$ excited by the seed pulse, and therefore seen by the long pulse, from equations \eqref{Eqn:WakePhi} and \eqref{Eqn:WakeDeltan}. Equation \eqref{Eqn:WakePhi} is solved using a fourth-order Runge-Kutta solver. The value of $\delta n/n_0$ is then assumed constant throughout the modulator stage and equation \eqref{Eqn:SpectralShiftSameFrequency} is used to compute the spectral shift.

\section{Particle-in-cell simulations}
Two-dimensional simulations were performed using the PIC code EPOCH (version 4.17.10) \cite{Arber:2015hc}.
Four simulations were performed. Modulation stage: (i) 600 mJ driver, and 50 mJ seed, both with a spot size of $w_0=\SI{30}{\mu m}$; and (ii) $600\times (50/30)^2=1670$ mJ driver, and $50\times (50/30)^2=140$ mJ seed, both with $w_0=\SI{50}{\mu m}$. Acceleration stage: (iii,iv) re-injection of pulse train generated from output of (i) and (ii), respectively into a plasma channel of matched spot size $w_M=\SI{30}{\mu m}$. For all simulations, the laser wavelength was $\lambda_0=\SI{1030}{nm}$ for both driver and seed. The seed and drive pulses were polarized out of and in the plane of the simulation, respectively. 

Ions were assumed to be  stationary for all the simulations presented in this work. Transverse Convolutional Perfectly Matched Layer (CPML) boundary conditions were tuned to ensure that light reaching the edge of the boundary box was absorbed. Open boundaries were used longitudinally. A second order Yee scheme for solving Maxwell’s equations was used to ensure high numerical stability and accurate dispersion relation. A longitudinal resolution $\Delta z=\lambda_0/50$ and transverse resolution $\Delta y=\lambda_0/2.5$ were used for simulations (i) and (ii). For simulation (iii) and (iv) the longitudinal resolution was increased to $\Delta z=\lambda_0/100$ to ensure accurate laser group velocities and corresponding electron dephasing lengths.

\subsection{Pulse compression}
The pulse trains used in PIC simulations (iii) and (iv) were generated from the laser field of the modulated drive pulses at the end of the modulator stage in PIC simulations (i) and (ii), respectively, by introducing GDD.

The laser field of the driver  following the modulation stage, $E^{\text{Mod}}_L(\xi,y)$, was Fourier transformed longitudinally to give, $\mathcal{E}_L^{\text{Mod}}(\omega,y)=\text{FFT}\left\{E^{\text{Mod}}_L(z,y)\right\}$. A constant second order spectral phase term $\psi_{\text{sim}}^{(2)}$ was introduced to yield a pulse train for the acceleration stage:
\begin{equation}
    E_L^{\text{Acc}}(z,y)=\text{IFFT}\left\{ \mathcal{E}_L^{\text{Mod}}(\omega,r)\exp{\left(i \frac{1}{2}(\omega-\omega_0)^2 \psi_{\text{sim}}^{(2)}\right)} \right\}.
    \label{eq:IFFTwithGDD}
\end{equation}
The value of $\psi_{\text{sim}}^{(2)}$ was optimized numerically to yield a pulse train with the highest peak intensity.
For simulations (i) and (ii), after a modulator stage of $L_{\text{mod}}=\SI{120}{mm}$, the optimal GDD was found to be $\psi_{\text{sim}}^{(2)}=\SI{-1480}{fs^2}$ and $\psi_{\text{sim}}^{(2)}=\SI{-1620}{fs^2}$ to generate pulse trains for injection into simulations (iii) and (iv), respectively.

\subsection{Pulse re-injection}
 The electromagnetic field components of the pulse train calculated above were converted into binary files and imported into a new EPOCH simulation as an external field in vacuum. This electromagnetic impulse causes forward and backward propagating light, where the energy division is not necessarily even. After running simulations in vacuum the data input is re-scaled to ensure that the laser propagating forward has the correct energy.

For simulation (iii) and (iv), it was necessary to couple the drive pulse train, of spot size $w_0^\mathrm{mod} = \SI{30}{\mu m}$ and $w_0^\mathrm{mod} = \SI{50}{\mu m}$, respectively, from vacuum into a plasma channel of matched spot size $w_M^\mathrm{acc} = \SI{30}{\mu m}$, with its entrance located at $z = 0$. This was achieved by applying the following procedure, for the drive pulse train located at $z<0$: (i) the transverse size of the field calculated from the simulation of the modulator and dispersive system was reduced by a factor $\eta=w_0^\mathrm{mod}/w(z)$; (ii) the field amplitude was increased by a factor $\eta^{-1}$; and (iii) the field was multiplied by the transverse phase term $\exp \left[i \theta(z,y)\right]$. Here, $w(z)=w_M^\mathrm{acc}\sqrt{1+z^2/z_R^2}$ is the spot size of a Gaussian beam of waist $w_0^\mathrm{acc}= w_M^\mathrm{acc}$, $z_R=\pi (w_M^\mathrm{acc})^2/\lambda_0$, and,

\begin{equation}
\theta(z,y)={\left(\frac{i ky^2}{2 R(z)}-i \varphi(z)\right)},
\end{equation}
where $R(z)=z+{z_{R}^{2}}/{z}$ and $\varphi(z)=\tan^{-1}\left(z/z_R\right)$.
The addition of this phase ensures that the pulse trains, starting from vacuum, will reach a focus at the entrance of the plasma waveguide. 

\subsection{Laser-plasma energy transfer efficiency}
\begin{figure}[tb]
     \centering
     \includegraphics[width=\columnwidth]{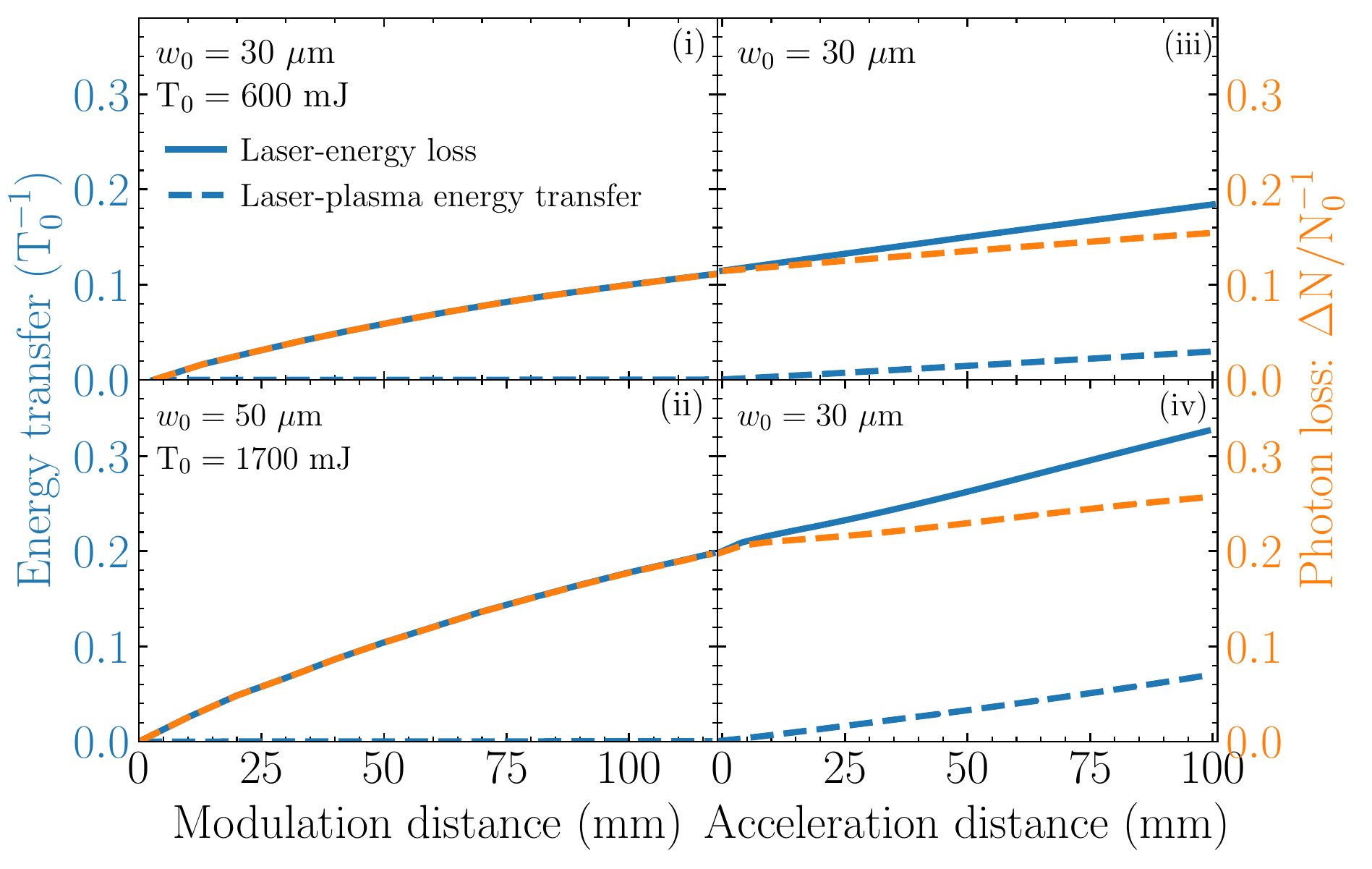}
     \caption{Calculated relative energy loss and photon number loss for the drive pulse in the modulator and accelerator sections for simulations (i-iv).}
     \label{fig:EnergyEfficiency}
 \end{figure}

The efficiency of the proposed scheme is estimated by comparing the relative energy loss of the drive-laser to the loss in the photon number. The relative energy loss of the drive laser $\Delta \text{T}(z)/\text{T}_0=(\text{T}_0-\text{T}(z))/\text{T}_0$, where $\text{T}_0$ is the initial energy of the drive laser at the start of the modulator stage and $\text{T}(z)$ is the remaining energy after propagating a distance $z$ in the modulator or acceleration stage. This is calculated from the PIC data using
\begin{align}
        \Delta \text{T}(z)/\text{T}_0=&1-\text{T}(z)/\text{T}_0 \nonumber\\ 
    =& 1-\frac{\sum_{y}\sum_{\xi}{|E_L(\xi,z,y)|^2}}{\sum_{y}\sum_{\xi}{|E_L(\xi,0,y)|^2}} \label{eq:EnergyLoss}
\end{align}
where $E_L$ is the electric field of the drive laser and $z=0$ is the start of the modulation stage. Similarly, the relative photon number loss $\Delta N(z)/N_0$, where $N_0$ is the initial photon number of the drive laser at the start of the modulator stage, is calculated in Fourier space:
\begin{align}
        \Delta N(z)/N_0 &= 1 - N(z)/N_0 \nonumber\\ =& 1-\frac{\sum_{y}\sum_{\omega}\left(|\mathcal{E}_L(\omega,z,y)|^2/{\omega}\right)}{\sum_{y}\sum_{\omega}\left(|\mathcal{E}_L(\omega,0,y)|^2/{\omega}\right)} \label{eq:PhotonNumber}
\end{align}
We make the assumption that leakage only reduces the photon number while leaving the laser spectrum unaltered, and that wake excitation does not alter the photon number but only affects the laser spectrum. In this way the energy loss can be written as $\Delta \text{T}/\text{T}_0=(\Delta \text{T}_\mathrm{plasma} + \Delta \text{T}_\mathrm{leak})/\text{T}_0$, as a sum of energy deposited into the plasma and energy lost through leakage, where $\Delta \text{T}_\mathrm{leak}/\text{T}_0=\Delta N /N_0$. The results of these calculations are shown in Fig.\ \ref{fig:EnergyEfficiency} for simulations (i-iv).

\begin{figure}[tb]
     \centering
     \includegraphics[width=\columnwidth]{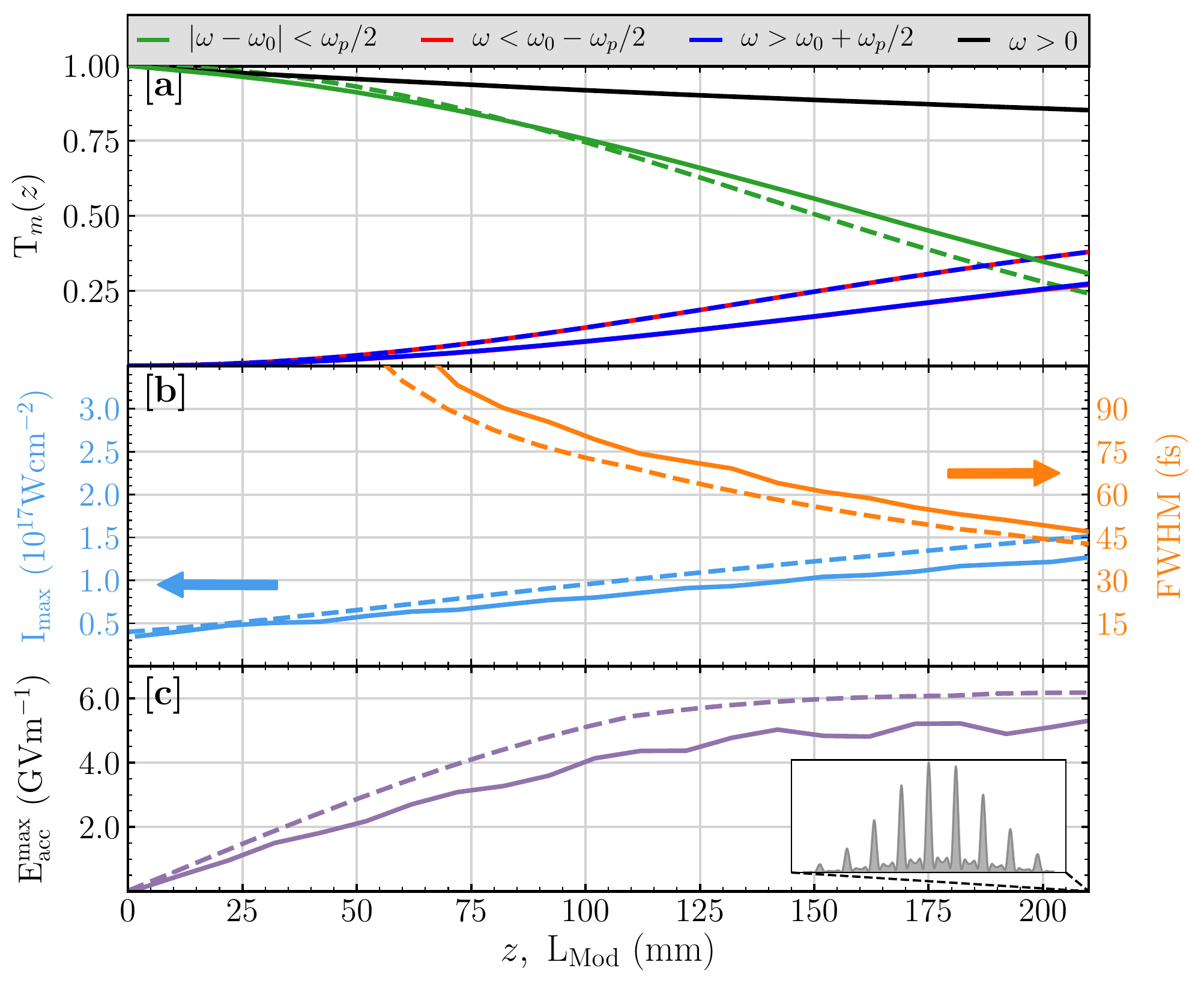}
     \caption{Comparison of the results of the 1D analytic model, equation \ref{Eqn:FullExpression}, (dashed lines) and PIC simulations (solid lines), plotted as  a function of the length of the modulator stage, for a \SI{1030}{nm} drive pulse and \SI{800}{nm} seed. All other parameters are as for Fig. 2 in the main paper. (a) The total transmitted energy (black) of the drive pulse, and the energies contained in the central band (i.e.\ $|\omega - \omega_0| < \omega_p /2$, green), and in the blue-shifted (i.e.\ $|\omega - \omega_0| > \omega_p /2$, blue) and red-shifted ($|\omega - \omega_0| < -\omega_p /2$, red) sidebands. (b) The peak intensity (blue), and FWHM duration (orange), of the most intense pulse in the train generated by applying a quadratic spectral phase, optimized to yield the highest-intensity pulse train, on the drive pulse emerging from the modulator stage. (c) Peak accelerating electric field produced by injecting into the accelerator stage the pulse trains generated at that point in the modulator.}
     \label{ModulationStage800nm}
 \end{figure}

\subsection{Driver and seed pulses of different wavelengths}

It is interesting to explore whether this scheme would work with drive and seed laser pulses of different wavelength, particularly since thin-disk or fibre lasers cannot easily generate sub-\SI{100}{fs} pulses. In this section we therefore investigate the use of a seed pulse provided by a  Ti:Sapphire laser operating at \SI{800}{nm} and a drive pulse from a thin-disk laser operating at \SI{1030}{nm}. This difference in wavelength will introduce a relative drift between the seed and drive pulses, and consequently, the red-shifting and blue-shifting phases of the seed-driven wake will drift with propagation relative to the drive pulse.

A PIC simulation with laser and plasma parameters identical to those of simulation (i) in the main paper, except for the use of an \SI{800}{nm}, \SI{50}{mJ}, seed pulse was performed. The drive pulse remains at \SI{1030}{nm}. The results are shown in Fig.\ \ref{ModulationStage800nm}; these may be compared with Fig.\ 3 in the main paper.

For this case, equation \eqref{Eqn:FullExpression} was solved numerically, taking into account the non-zero drift velocity $\Delta v =v_g(\SI{1030}{nm}) - v_g(\SI{800}{nm})$, to get the spectral shift of the drive pulse for different modulator lengths, as shown in figure \ref{ModulationStage800nm}(a). Following optimization for the highest intensity pulse train, as described above, the intensity and pulse duration of the highest-intensity pulse is shown in figure \ref{ModulationStage800nm}(b) for the PIC data and linear propagation theory. The on-axis accelerating electric wakefield driven by these pulse trains is estimated using equation \eqref{Eqn:WakePhi} and shown in figure \ref{ModulationStage800nm}(c) for PIC data and the 1D analytic model. This shows that pulse trains capable of driving a wakefield at $\sim \SI{5}{GV/m}$ -- the same maximum amplitude reached by the pulse trains in Fig.\ 3 in the main paper -- can be produced by compression after a modulator length of \SI{150}{mm}, in comparison to the \SI{100}{mm} required for pulses of the same wavelength.

It can be seen that, for these parameters, the drive pulse can still be modulated and compressed to form a train of short pulses, although the drift between the seed and drive pulses causes a longer modulator to be necessary. We note that the amplitude of the wakefield is the same as when the seed and drive pulses have the same wavelength. The reason for this is that the amplitude of the plasma wave excited by a train of short pulses depends only on the total laser energy of the train, provided that the duration of each pulse is sufficiently short. By comparing Fig.\ \ref{ModulationStage800nm} with Fig.\ 3 in the main paper we see that  saturation of the amplitude of the plasma wave is reached for pulse durations less than approximately 45 fs, which is short compared to the 220 fs plasma period.

\subsection{Validity of 2D simulations}
\begin{figure}[tb]
     \centering
     \includegraphics[width=\columnwidth]{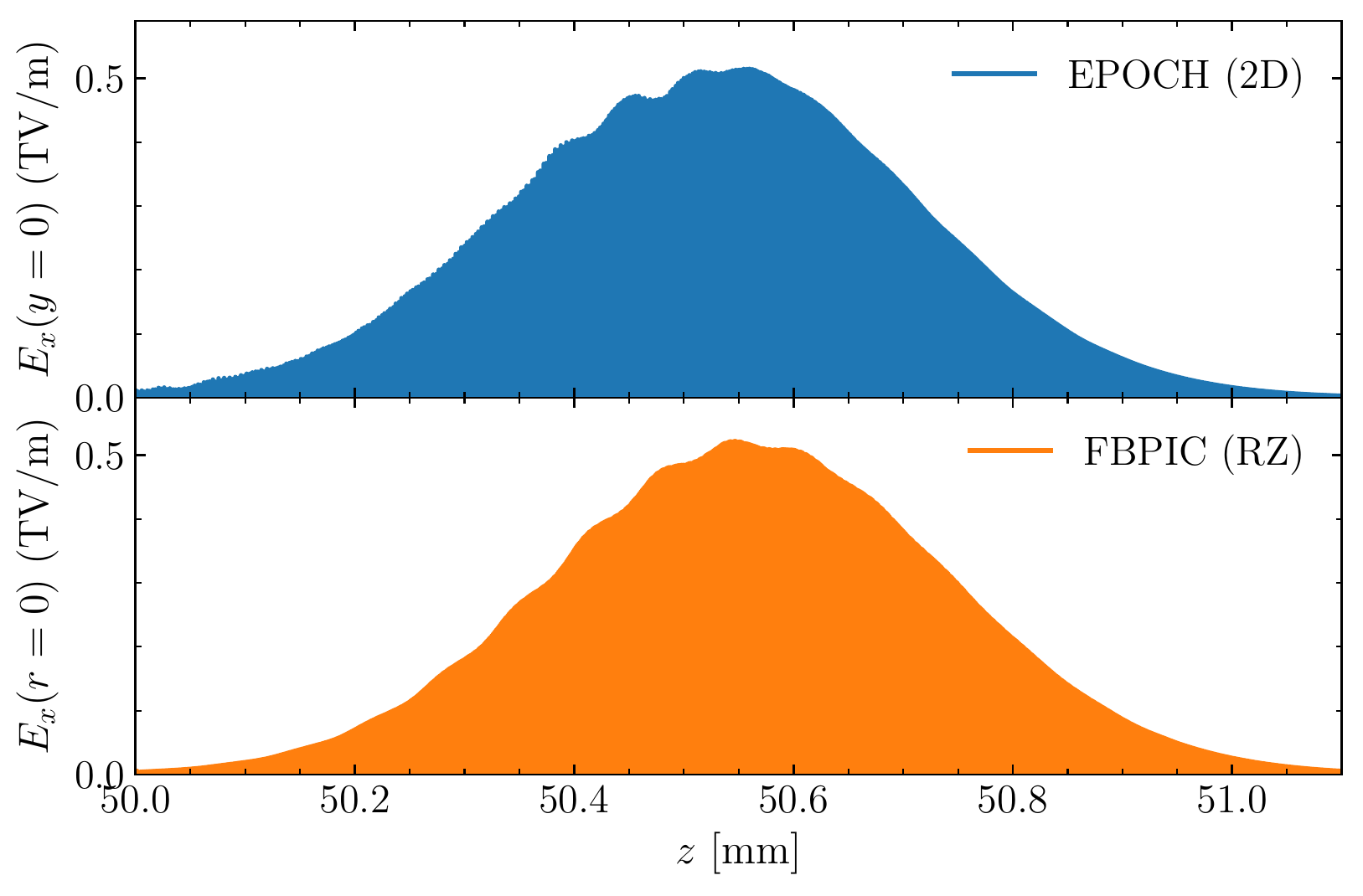}
     \caption{Comparison of the modulus of the laser field of the \SI{1.7}{J}, $\SI{50}{\micro m}$ drive pulse in simulation (ii) of the main paper, after \SI{50}{mm} of co-propagation with the seed wake, for the 2D EPOCH simulation in the main manuscript and the same simulation using FBPIC in cylindrical symmetry.  }
     \label{EPOCH_FBPIC_Figure4}
 \end{figure}

To ensure that the 2D EPOCH simulations captured the relevant laser-plasma interactions, i.e that the long pulse propagates through the plasma waveguide without significant self-modulation, initial tests were performed using the code FBPIC, which employs a cylindrical PIC algorithm with azimuthal Fourier decomposition \cite{Lehe2016}. Figure \ref{EPOCH_FBPIC_Figure4} compares the drive pulse in simulation (ii), after \SI{50}{mm} co-propagation with the seed wake in the plasma waveguide, in a 2D EPOCH simulation and a FBPIC simulation using 2 azimuthal modes. The good agreement between these codes shows that, for the parameters of interest, the 2D simulations in the main manuscript capture the necessary laser-plasma interactions.

\end{document}